\documentclass[final]{ws-procs9x6}
\usepackage{graphicx}

\def\lab#1      {\hbox{\small #1} }

\newcommand{\be}{\begin{eqnarray}}
\newcommand{\ee}{\end{eqnarray}}
\newcommand{\ben}{\begin{eqnarray*}}
\newcommand{\een}{\end{eqnarray*}}
\newcommand{\la}{\langle}
\newcommand{\ra}{\rangle}
\newcommand{\half}{\frac{1}{2}}

\def\mb#1         {\mbox{\boldmath $#1$}}
\def\diffn#1	  {\Delta^{-}_{#1}}

\begin{document}

\title{Consistent Definitions of Flux and Electric and Magnetic Current in Abelian Projected  $SU(2)$
Lattice Gauge Theory}

\author{Richard W. Haymaker\footnote{\uppercase{T}alk presented
at \uppercase{C}onfinement 2003, \uppercase{RIKEN}, 
\uppercase{J}uly  21 - 24, 2003 by \uppercase{R}. \uppercase{H}aymaker}
~and 
\uppercase{T}akayuki \uppercase{M}atsuki\footnote{\uppercase{P}ermanent Address: \uppercase{T}okyo \uppercase{K}asei 
\uppercase{U}niversity, 1-18-1 \uppercase{K}aga, \uppercase{I}tabashi, \uppercase{T}okyo 173-8602, 
\uppercase{J}apan}}
\address{Department of Physics and Astronomy, 
 Louisiana State University, \\
Baton Rouge, Louisiana 70803-4001, USA\\
E-mail: haymaker@rouge.phys.lsu.edu,matsuki@tokyo-kasei.ac.jp}

\maketitle

\abstracts{
Through the use of a lattice $U(1)$ Ward-Takahashi identity,  one can find a precise definition of flux and
electric four-current that does not rely on the continuum limit.  The magnetic four-current defined for example 
by the DeGrand-Toussaint construction introduces order $a^2$ errors in the field distributions.   We advocate 
using a single definition of flux in order to be consistent with both the electric and magnetic 
Maxwell's equations at any lattice spacing. In a $U(1)$ theory the monopoles are slightly smeared by this 
choice, i.e. are no longer associated with a single lattice cube.  In Abelian projected $SU(2)$ the consistent 
definition suggests further modifications.   For simulations in the scaling window, we do not foresee 
large changes in the  standard analysis of the dual Abrikosov vortex
in the maximal Abelian gauge because the order $a^2$ corrections have small fluctuations and tend to
cancel out.  However in other gauges, the consequences of our definitions could lead to large effects 
which may help in understanding the choice of gauge.   We also examine the effect of truncating all 
monopoles except for the dominant cluster on the profile of the dual Abrikosov vortex. 
}

\section{Introduction}
\label{Introduction}

Dual superconductivity has long been suggested as a possible mechanism for quark
confinement signaled by a spontaneously broken $U(1)$  gauge symmetry and 
manifested by a dual Abrikosov vortex between quark and anti quark.
This picture was verified  some time ago in Abelian projected SU(2) lattice gauge theory 
in the maximal Abelian gauge\cite{sbh}.   More recently further studies have 
elaborated this  picture\cite{bss,gips,koma}.   

As in all lattice calculations, there is freedom in choosing lattice operators,
requiring only that they agree with the continuum definition to lowest
order in the lattice spacing $a$.  However we have the opportunity in these
studies to be more precise by incorporating the lattice Ward-Takahashi identity
derived from the residual $U(1)$ gauge symmetry\cite{dhh}. 
This gives an Ehrenfest relation for the expectation 
value of the fields and currents giving  the electric Maxwell equations 
exactly at finite lattice spacing.  Interestingly, this defines a unique lattice expression 
for the field strength or flux and the electric and magnetic currents.  

In the present work, we  examine the impact of this
on the study of the dual Abrikosov vortex.
To be consistent, the magnetic Maxwell equation must use the same definition as the electric one.  
However the standard 
DeGrand-Toussaint\cite{dt} [DT] definition of the magnetic current is based on a 
different definition of flux, resulting in inconsistencies in the  magnetic Maxwell equation.
We argue here that one should alter the DT construction, using a single definition
of flux throughout.  A consequence is that the magnetic current no longer contains
discrete monopoles but rather a more general magnetic charge distribution.  In effect,
the monopoles are smeared in our picture.

This consistency question is only relevant at finite lattice spacing and all these 
concerns go away in the continuum limit.   Nevertheless
it is desirable to have a consistent treatment of total flux in the vortex determined
from the electric Maxwell equation, and the profile shape determined by the magnetic
Maxwell eqution at fixed lattice spacing.   Further we note that 
the finite lattice spacing effects 
are significant for the values of $\beta$ that we often use for calculations.

We also report on the effect of truncating DT monopole loops, keeping only the one large
connected cluster.  This truncation is expected to have no effect on the confinement
signal\cite{ht,z}.  This should manifest itself here in that the tail of the profile
of the  magnetic current of the vortex are unaffected by the truncation.
We find this to be the case.
This procedure requires that the magnetic current consists of discrete monopoles.  Truncation
is not well defined in the above smeared monopole picture.  Hence we take the conventional
view in presenting these results that the continuum limit is required to obtain the 
consistency described above.

\section{Three definitions of flux}

Let us consider three definitions of field strength, all agreeing to lowest order in $a$.
The first definition was used by DT to define 
monopoles:
\ben
  \widehat{F}_{\mu \nu}^{(1)}&=&\theta_\mu(\mb{m} )-\theta_\mu(\mb{m} +\nu)
  - \theta_\nu(\mb{m} ) + \theta_\nu(\mb{m} +\mu) - 2 \pi n_{\mu\nu},\\
  &\equiv& \theta_{\mu \nu}(\mb{m} ) - 2 \pi n_{\mu \nu},
\een
where $\theta_\mu(\mb{m} )$ refers to the $U(1)$ link angle in the domain
$-\pi < \theta_\mu < +\pi$.  The integers $n_{\mu \nu}$
are determined by requiring that  $-\pi < \widehat{F}_{\mu \nu}^{(1)} < +\pi$.
That is $\widehat{F}_{\mu \nu}^{(1)}$ is a periodic function of $\theta_{\mu \nu}$
with period $2 \pi$.   Here quantities
with $~\widehat{ }~$ mean those which appear in the lattice numerical calculation
without appending factors of $e$ and $a$.

The second and third definitions  gives the exact electric Maxwell equation for lattice averages
\be
\Delta_\mu^{-}
\la  \widehat{F}_{\mu\nu}^{(i)}\ra_W &=&
 \la \widehat{J}_\mu^{e(i)} \ra_W, \quad \quad i = 2,3,
\label{judy}
\ee
where 
\ben
\la \cdots \ra_W = \frac{\la \cdots e^{i \theta_W}\ra}{\la e^{i \theta_W}\ra},
\een
for the cases of $U(1)$ gauge theory and $SU(2)$ gauge theory respectively.  We give the  $U(1)$ derivation in 
detail since it is straight forward and contains the essential points of the argument.  
There are significant complications in the $SU(2)$ case and hence we just sketch that derivation.

\subsection{Flux in the U(1) gauge theory}

 Consider
\ben
Z_W(\epsilon_\mu(\mb{m} )) = \int [d \theta] \sin \theta_W
\exp\left( \beta S,
\right)
\een
\ben
S = \sum_{n,\;\mu > \nu} 
\cos \theta_{\mu \nu}(\mb{n} ),\;\;\;\;\;\;\;\;\;\;   \beta = \frac{1}{e^2}.
\een
The subscript of $Z_W(\epsilon_\mu(\mb{m} )) $ refers to the incorporation of the source into the partition function
and the argument is a variable defined as the shift of one particular link,
$\theta_\mu (\mb{m} ) \rightarrow \theta_\mu(\mb{m} ) +  \epsilon_\mu(\mb{m} ) $.
This translation
can be transformed away since the measure is invariant under such an operation.   
Therefore   
\be
0 = \delta Z_W 
&=&
\left. 
\int [d\theta]\sin\theta_W \exp(\beta S)
\right|_{\theta_\mu \rightarrow \theta_\mu+\epsilon_\mu}
- 
\int [d\theta]\sin\theta_W \exp(\beta S)\nonumber
\\
&=&
\epsilon_\mu
\int [d \theta]\left(  \delta_\mu(\mb{m} ) \cos \theta_W  + 
\sin \theta_W \frac{1}{e^2}\frac{\partial S}{\partial \theta_\mu}\right)
\exp\left( \beta S \right) ,
\label{charlie}
\ee
where $\delta_\mu(\mb{m} ) = \pm 1$  if $ \mb{m} $ labels a $\pm$ directed 
link and $ = 0 $ otherwise. This is the static current generated by the 
Wilson loop. 
\be
\delta_\mu(\mb{m} ) = \widehat{J}^e_\mu
\label{maggie}
\ee

Next evaluate the derivative of $S$.  Isolate the six plaquettes 
affected by the shift
\ben
&&S =
\sum_{\nu \ne \mu} 
\left[
\cos 
\left\{
\theta_\mu(\mb{m} ) + 
\theta_\nu(\mb{m} + \mu)- 
\theta_\mu(\mb{m} + \nu)- 
\theta_\nu(\mb{m} ) 
\right\}  
\right. \\
&&
\quad \quad \quad \quad 
+\left.
\cos 
\left\{
\theta_\mu(\mb{m} ) -
\theta_\nu(\mb{m} + \mu - \nu) - 
\theta_\mu(\mb{m} - \nu)+ 
\theta_\nu(\mb{m} - \nu) 
\right\}  
\right] + \cdots,
\een
\ben
&& \frac{\partial S}{\partial \epsilon_\mu(\mb{m} )} = 
\sum_{\nu \ne \mu} 
\left[
-\sin 
\left\{
\theta_\mu(\mb{m} ) + 
\theta_\nu(\mb{m} + \mu)- 
\theta_\mu(\mb{m} + \nu)- 
\theta_\nu(\mb{m} ) 
\right\}  
\right. \\
&&
\quad \quad \quad \quad 
-\left.
\sin 
\left\{
\theta_\mu(\mb{m} ) -
\theta_\nu(\mb{m} + \mu - \nu) - 
\theta_\mu(\mb{m} - \nu)+ 
\theta_\nu(\mb{m} - \nu) 
\right\}  
\right],
\een
\be
\frac{\partial S}{\partial \epsilon_\mu(\mb{m} )} 
&=& - \Delta_{\nu}^{-} \sin \theta_{\mu \nu}(\mb{m} ).
\label{betty}
\ee
Using  Eqn.(\ref{judy}) and Eqn.(\ref{betty}) we can see that Eqn.(\ref{charlie}) 
is the form of Maxwell equations for 
averages.  
\be
\frac{1}{e^2}\Delta_{\nu}^{-}
\la \widehat{F}^{(2)}_{\mu \nu} \ra_W
&=&
 \widehat{J}^{e}_\mu(\mb{m} ),
\label{john}
\ee
where
\ben
\la \cdots \ra_W &=& \frac{\la \sin \theta_W \cdots \ra}{\la\cos \theta_W \ra}.
\een
Since the charged line in a Wilson loop is closed the electric current is conserved.  The local 
statement of conservation  is
\ben
0 = 
\Delta^-_\mu
\Delta^-_\nu
\la \widehat{F}^{(2)}_{\mu \nu}\ra_W
=
\Delta^-_\mu
\widehat{J}^{e}_\mu . 
\een
It is straightforward to verify on the lattice that the LHS of Eqn.(\ref{john}) gives $-1,  0 , 1$ 
depending on its position
with respect to the Wilson loop.  An alternative definition such as  $\widehat{F}^{(1)}_{\mu \nu}$ 
need not vanish off the Wilson loop  nor give $\pm 1$ on the Wilson loop and hence would introduce an error.  
In a $U(1)$ theory there is no dynamical charge density, all charge resides on the Wilson loop.

\subsection{U(1) flux in the SU(2) theory in the maximal Abelian gauge}

We restrict our attention to the maximal Abelian gauge defined as a  local maximum of
\ben
R &=&  \sum_{n,\mu}\lab{tr} \left\{\sigma_3 U_\mu(n) \sigma_3 U^{\dagger}_\mu(n)  \right\},
\een
over the set of gauge transformations $\left\{g(m) = e^{i \alpha_i (m) \sigma_i} \right\}$, 
$U \longrightarrow U^g$.  Taking $U$ to be the stationary value, the stationary condition is given by
\ben
F_{j n}[U] &=& \left. \frac{\partial R[U^g]}{\partial \alpha_j(n)}\right|_{\alpha = 0} = 0.
\een
The second derivatives entering in the Jacobian are given by
\ben
M_{j n; i m}(U) &=& \left. 
\frac{\partial^2 R[U^g]}{\partial \alpha_j(n)\partial \alpha_i(m)}\right|_{\alpha = 0} .
\een
The partition function is
\be
Z_W^{g.f.} (\epsilon_{\mu}^3(\mb{m} )) &=& \int [dU] \; 
\half Tr[ i\sigma_3 U_W (\mb{n} ) ] \;
\exp\left( \beta S \right)
\prod_{j n}\delta(F_{j n}[U]) \; \Delta_{FP}, 
\label{su2-3}
\ee
where the Faddeev-Popov  Jacobian  is
\ben
\Delta_{FP} &=&  \lab{det} | M_{j n; i m} (U) |.
\een
An infinitesimal shift in this partition function has the added complication that it violates the
gauge condition.  This can be corrected by an infinitesimal accompanying gauge transformation.  Thus
the shift in one link affects all links.  However experience has shown that the effect drops off
rapidly with distance from the shifted link.  

The derivative of the partition function Eqn.(\ref{su2-3})with respect to $\epsilon_{\mu}^3(\mb{m} ) $ gives
\be
0 &=& 
\int [dU]  \left\{\delta_\mu(\mb{m} )\half Tr[U_W (\mb{n} ) ]   \right\}
\exp\left( 
\beta S
\right)
\prod_{j n}\delta(F_{j n}[U])  \Delta_{FP} 
\nonumber
\\
&+& 
\int [dU]  \left\{ 
\beta \half Tr[ i\sigma_3 U_W (\mb{n} ) ] 
\frac{\partial S}{\partial \epsilon_{\mu}^3(\mb{m} )}   \right\}
\exp\left(
\beta S
\right)
\prod_{j n}\delta(F_{j n}[U])  \Delta_{FP} 
\nonumber
\\
&+&
 \int [dU] 
\half Tr[ i\sigma_3 U_W (\mb{n} ) ] 
\exp\left( \beta S \right)
\frac{\partial }{\partial \epsilon_{\mu}^3(\mb{m} )}
\left\{\prod_{j n}\delta(F_{j n}[U])  \Delta_{FP}\right\}.
\label{su2-4}
\ee
The third integral contains terms in the Ward Takahasni identity coming from the gauge fixing including ghost contributions.

We can cast this into the form of the electric Maxwell's equations for averages as in the case of the 
U(1) theory.  However there are now more terms in the current.
Consider the standard U(1) parametrization of an SU(2) element:
\ben
U_{\mu}(\mb{n} ) 
&=& 
\left( 
\begin{array}{cc}
C_{\mu} (\mb{n} ) e^{i \theta_{\mu} (\mb{n} )}  
& S_{\mu} (\mb{n} )e^{i \left\{\gamma_{\mu} (\mb{n} ) - \theta_{\mu} (\mb{n} ) \right\}}\\
- S_{\mu} (\mb{n} )e^{-i \left\{\gamma_{\mu} (\mb{n} ) - \theta_{\mu} (\mb{n} ) \right\}}
&
C_{\mu} (\mb{n} ) e^{-i \theta_{\mu} (\mb{n} )}
\end{array}
\right),
\\
&=& 
\left( 
\begin{array}{cc}
C_{\mu}  & S_{\mu} e^{i \gamma_{\mu} } \\
- S_{\mu} e^{-i \gamma_{\mu} }  & C_{\mu}  \\
\end{array}
\right)
\left( 
\begin{array}{cc}
e^{i \theta_{\mu} } & 0 \\
0& e^{-i \theta_{\mu} } \\
\end{array}
\right),
\een
where
\be
C_{\mu} (\mb{n} ) &\equiv& \cos \phi_{\mu} (\mb{n} ),
\nonumber
\\
S_{\mu} (\mb{n} ) &\equiv& \sin \phi_{\mu} (\mb{n} ). 
\label{su2-5}
\ee
In the Abelian projection
factored form, the righthand factor contains the $U(1)$ photon, parameterized by $\theta$.  The
 lefthand factor contains  the charged coset matter fields, parameterized  by $\phi$ and $\gamma$. 
  The transformation properties are well known and reviewed in DiCecio et.al.\cite{dhh}.

We consider an alternative separation into diagonal and off-diagonal parts which is needed in defining the
flux.
\ben
U_{\mu}(\mb{n} ) 
&=& 
\left( 
\begin{array}{cc}
C_{\mu}  e^{i \theta_{\mu} }  
& 0\\
0&
C_{\mu}  e^{-i \theta_{\mu} }
\end{array}
\right)
+
\left( 
\begin{array}{cc}
0& S_{\mu} e^{i \left(\gamma_{\mu}  - \theta_{\mu}  \right)}\\
- S_{\mu} e^{-i \left(\gamma_{\mu}  - \theta_{\mu}  \right)}
&
0
\end{array}
\right),
\\\\
&=& D_{\mu}(\mb{n} ) + O_{\mu}(\mb{n} ).
\een
The off-diagonal part is the charged matter field 
$\Phi_{\mu} \equiv S_{\mu} e^{-i \left(\gamma_{\mu} - \theta_{\mu} \right)}$.
and diagonal part includes the photon,  $e^{i \theta_{\mu} }$, but also a neutral remnant of the matter field
$\sqrt{1 - |\Phi_\mu|^2}$ which   $\rightarrow 1$  in the limit $a \rightarrow 0 $.

To cast Eqn.(\ref{su2-4}) into the form of a current conservation law, 
we first consider the terms to zeroth order in $O_\mu$.
First the Wilson loop.  Isolating the diagonal contributions gives
\ben
 U_W  &=&  D_W  +  \widetilde{U}_W,
\een
where $D_W$ is the product of the diagonal parts 
\ben
 \half Tr[  D_W (\mb{n} ) ] &=&   \prod_W  C_\mu(\mb{n})\cdot \cos \theta_W, \\
 \half Tr[ i\sigma_3 D_W (\mb{n} ) ] &=&  - \prod_W  C_\mu(\mb{n})\cdot \sin \theta_W .
\een
We adhere to the standard  choice of  an Abelian Wilson loop in which we drop any contributions due to 
off diagonal elements $O_\mu$ and further take the factors $ C_\mu(\mb{n}) = 1$ giving
\ben
\half Tr[U_W^{\lab{Abelian} } (\mb{n} )] &=& \cos \theta_W, \\
\half Tr[i\sigma_3U_W^{\lab{Abelian} } (\mb{n} )] &=& -\sin \theta_W.
\een

Second, consider the action. Write
\ben
S =  \sum_{n,\;\mu > \nu} 
 \half\lab{Tr} [D_{\mu \nu}(\mb{n} )]  + \widetilde{S},
\een
where $\widetilde{S}$ contains terms involving $O_{\mu}(\mb{n} )$.
\ben
&& \frac{\partial (S - \widetilde{S})}{\partial \epsilon_\mu^3(\mb{m} )} = 
\sum_{\nu \ne \mu} 
\left[
\half \lab{Tr}
\left\{ -i\sigma_3(\mb{m} )
D_\mu(\mb{m} ) 
D_\nu(\mb{m} + \mu)
D_\mu^\dagger(\mb{m} + \nu) 
D_\nu^\dagger(\mb{m} ) 
\right\}  
\right. \\
&&
\quad \quad \quad \quad 
+\left.
\half \lab{Tr}
\left\{ -i \sigma_3(\mb{m} )
D_\mu(\mb{m} ) 
D_\nu^\dagger(\mb{m} + \mu - \nu) 
D_\mu^\dagger(\mb{m} - \nu)
D_\nu(\mb{m} - \nu) 
\right\}  
\right].
\een
Since all matrices are diagonal we can simplify:
\ben
\frac{\partial (S - \widetilde{S})}{\partial \epsilon_\mu^3(\mb{m} )} &=& 
\sum_{\nu \ne \mu} 
\Delta^-_\nu 
\left[
C_\mu(\mb{m} ) 
C_\nu(\mb{m} + \mu)
C_\mu(\mb{m} + \nu) 
C_\nu(\mb{m} ) 
  \sin \theta_{\mu \nu}(\mb{m} )
\right] .
\een
The quantity in square brackets is antisymmetric in $\mu \nu$ and we identify this as proportional
to the field tensor.
\ben
\widehat{F}^{(3)}_{\mu \nu} &\equiv& 
C_\mu(\mb{m} ) 
C_\nu(\mb{m} + \mu)
C_\mu(\mb{m} + \nu) 
C_\nu(\mb{m} ) 
  \sin \theta_{\mu \nu}(\mb{m} )
\een

Returning to the identity, Eqn.(\ref{su2-4}), and using the notation
\ben
 \int [dU]  \{ \cdots \} \exp\left( \beta S \right)
\prod_{j n}\delta(F_{j n}[U]) \Delta_{FP}  &=& 
 \la \cdots \ra_{g.f.}
\een
we obtain
\ben
0 &=& 
 \delta_\mu(\mb{m} )\la \cos \theta_W \ra_{g.f.} 
 - \beta  \la \sin \theta_W  \Delta^-_\nu  \widehat{F}^{(3)}_{\mu \nu}\ra_{g.f.} 
- \beta \la \sin \theta_W \frac{\partial \widetilde{S}}{\partial \epsilon_\mu^3(\mb{m} )} 
\ra_{g.f.} \\&&
+ \lab{ gauge fixing terms + ghost terms }.
\een
Rearranging terms as in the $U(1)$ case, we get
\ben
\beta  
 \frac{\la \sin \theta_W  \Delta^-_\nu  \widehat{F}^{(3)}_{\mu \nu}\ra_{g.f.} }{\la \cos \theta_W \ra_{g.f.} }
 = 
 \delta_\mu(\mb{m} )
 - \beta \frac{ \la \sin \theta_W \frac{\partial \widetilde{S}}{\partial \epsilon_\mu^3(\mb{m} )} 
\ra_{g.f.}}{\la \cos \theta_W \ra_{g.f.} }
+ \frac{\lab{g.f. \& ghosts} }{\la \cos \theta_W \ra_{g.f.} },
\een
\be
\Delta^-_\nu 
 \la    F^{(3)}_{\mu \nu}\ra_{W,g.f.}
&=&
 \widehat{J}_\mu^{\lab{(e) Abelian Wilson loop} } + 
\widehat{J}\mu^{\lab{(e) matter fields} } + \nonumber
\\&&
\widehat{J}\mu^{\lab{(e) gauge fixing} } + 
\widehat{J}\mu^{\lab{(e) ghosts} } .
\label{jed}
\ee
The `Abelian Wilson loop' term is analogous to the above $U(1)$ case.  The `charged matter field' term arises 
from the off-diagonal elements of links in the action expression and would contribute 
without gauge fixing.  The `gauge fixing' term arises
from the corrective gauge tranformation that accompanies the shift of a link.  The `ghost' term arises
from  the shift and accompanying gauge transformation on the Faddeev-Popov determinant.  See DiCecio 
et.al.\cite{dhh} for a complete derivaton of all terms.  

It is important here to note that we are not interested in distinguishing the various 
contributions to the current in the present work.  We are only interested in the total
current and that can be obtained from the LHS of Eqn.(\ref{jed}).

\section{Consistency with the magnetic Maxwell equation}

Having defined a unique flux $\widehat{F}_{\mu \nu}^{(i)}$ through the electric Maxwell equation, the magnetic
Maxwell equation is
\ben
 \frac{1}{2}\epsilon_{\mu\nu\rho\sigma}\Delta_\nu^{+}
  \widehat{F}_{\rho\sigma}^{(i)} &=&
  \widehat{J}_\mu^{m(i)} \quad \quad i = 2,3.
\een
However the standard DT definition of current is 
\ben
  \widehat{J}_\mu^{m(1)} &=&
 \frac{1}{2}\epsilon_{\mu\nu\rho\sigma}\Delta_\nu^{+}
  \widehat{F}_{\rho\sigma}^{(1)} .
\een
and hence if we use the conventional  $\widehat{F}^{(1)}$ to define the monopole current, and
$\widehat{F}^{(2)}$ or $\widehat{F}^{(3)}$ respectively for $U(1)$ and $SU(2)$ theories
to get an exact expression for flux in the confining string, then the magnetic Maxwell equation 
is violated.  

The solution is to relax the requirement that we use the DT monopole definition and use
$\widehat{F}^{(2)}$ or $\widehat{F}^{(3)}$ instead when defining monopoles. A simple configuration
for the $U(1)$ case ($\widehat{F}^{(2)}$) will illustrate the effect.  Consider a single 
DT monopole with equal flux out of the six faces of the cube.  Then the ratio of the 
$\widehat{F}^{(2)}$ flux out of this
cube compared to the DT $\widehat{F}^{(1)}$ flux gives
\ben
\frac{6 \sin (2 \pi/6)}{6 (2 \pi/6)} \approx 0.83.
\een
Since the current is conserved, the balance is made up by magnetic charge in the neighboring cells.
On a large surface the total flux is the same for the two definitions.

In Fig.(1) we plot the $\widehat{F}_{\mu \nu}^{(1)}$
as a function of $\theta_{\mu \nu}$, giving a ``sawtooth" shape.
Monopoles occur as a consequence of $\widehat{F}^{(1)}$  crossing the
sawtooth edge, giving a mismatch of $2 \pi$ in the flux out of a cube.  
The sine function has no such discreteness and so the 
notion of discrete Dirac strings and Dirac monopoles is modified.  However
as one approaches the continuum limit, the action will drive the plaquette
to zero and then the regions where the sawtooth differs from the sine
are suppressed.  Hence we expect both forms to give the standard
Dirac picture in the continuum limit.



\label{venus}
\includegraphics[trim=0 50mm 0 40mm,scale=0.35]{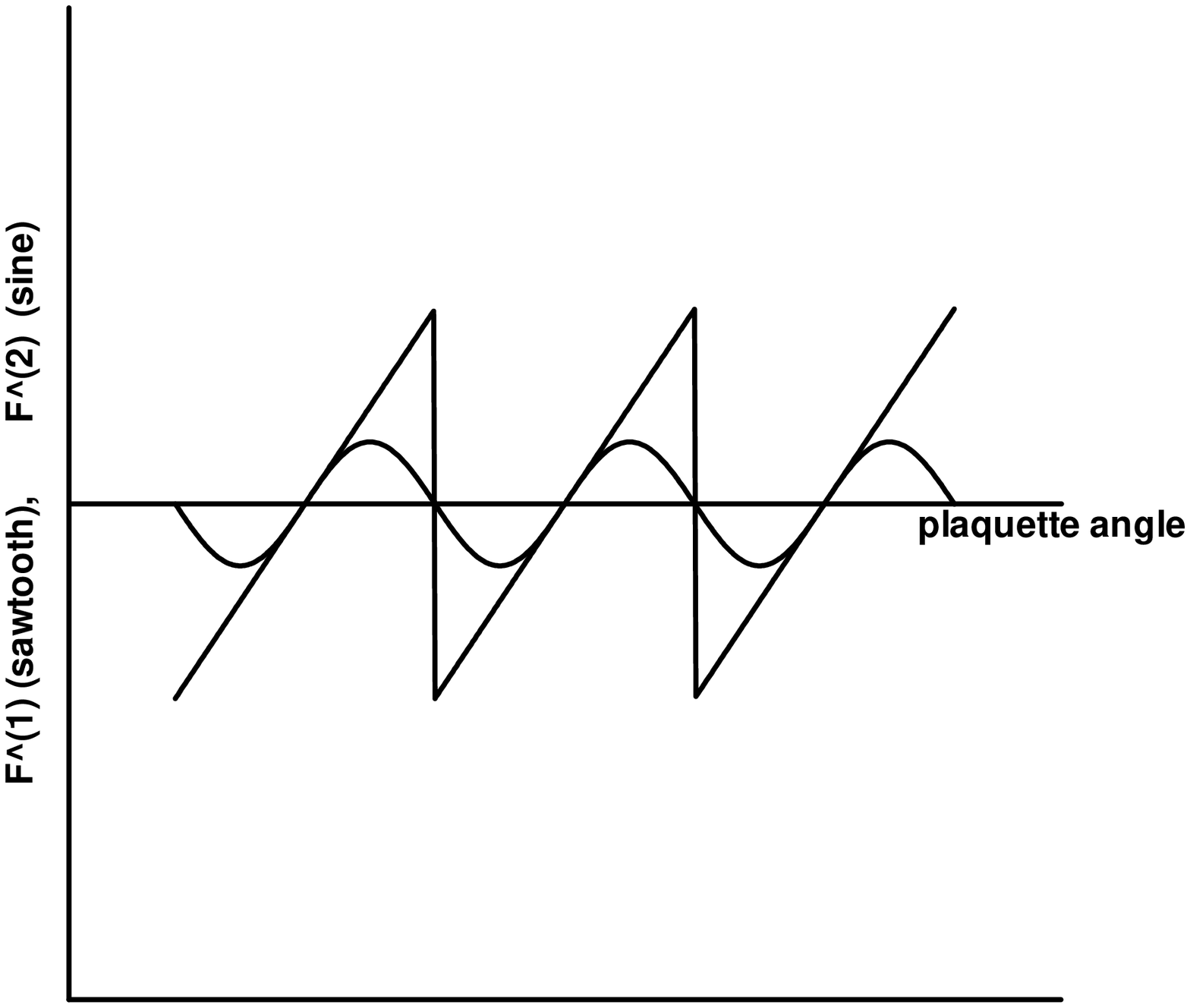}

\begin{flushleft}
Figure 1. $\widehat{F}_{\mu \nu}^{(1)}$ (sawtooth) and
$\widehat{F}_{\mu \nu}^{(2)}$ (sine)
as a function of the plaquette angle $\theta_{\mu \nu}$
\end{flushleft}
\vspace{0.5cm}

The electric Maxwell equation determines the total flux in the confining string and the 
magnetic Maxwell equation determines the 
transverse shape.  Further the latter enters directly in the determination
of the London penetration length, $\lambda_d$.  To see this let us consider the classical 
Higgs theory which we use to model the simulation data.  The dual field is given by
\ben
\widehat{G}_{\mu \nu}(\mb{m} )  &=&
 \Delta_\mu^{-} \theta_\nu^{(d)}(\mb{m} ) - \Delta_\nu^{-} \theta_\mu^{(d)}(\mb{m} ),
\een
where $\theta_\mu^{(d)}(\mb{m} )$ is a dual link variable.  
Let us choose to break the gauge symmetry spontaneously 
through  a constrained Higgs field.
\ben
\Phi(\mb{m} ) = v \rho(\mb{m} ) e^{i\chi(\mb{m} )}, \quad \quad \rho(\mb{m} )=1.
\een
Under these conditions the magnetic current simplifies to
\ben
  \widehat{J}^m_\mu (\mb{m} ) = 2e_m^2 v^2
  \sin\left\{
  \theta_\mu(\mb{m} )+ 
\chi_\mu(\mb{m} + \mu) -\chi_\mu(\mb{m} ) 
  \right\}.  
\een
where $e_m$ is the magnetic coupling. The magnetic Maxwell equation is
\be
  \Delta_\mu^+ \widehat{G}_{\mu \nu} &=& \widehat{J}_\nu^m .
\label{fred}
\ee

For small $\theta^{(d)}$ it is easy to see that there is a London relation of the form
\be
  \widehat{G}_{\mu \nu}(\mb{m} ) 
  = \frac{1}{2e_m^2 v^2}\left(\Delta^-_\mu \widehat{J}^m_\nu (\mb{m} ) - 
  \Delta^-_\nu \widehat{J}^m_\mu (\mb{m} )\right).
\label{bob}
\ee
Taking the confining string along the $3$ axis and choosing $\mu = 1$ and $\nu = 2$ 
we see that the profile of the third component of curl of the magnetic current 
must match the third component of the electric flux profile. 
This assumes an infinite Higgs mass $M_H$. With a finite mass 
there is a transition region of size $\sim 1/M_H$ in the core of the vortex but the
above London relation must hold sufficiently far outside the core. 

Combining Eqns.(\ref{fred}) and (\ref{bob}) we get the relation
\ben
  \left(1-\lambda_d^2
\Delta_\mu^+ \Delta_\mu^- \right) \left<\widehat{E}_3(\mb{m} )\right>_W =0,
\lambda_d^{-1} =  e_m v\sqrt{2}
\label{mary}
\een
{\em The corresponding equations in the simulation of the 
$SU(2)$ theory must also be satisfied in order to arrive 
at this correct expression for $\lambda_d$ and hence the importance of our 
definitions.}  

\section{Numerical Results}

We generated $208$ configurations  on a $32^4$ lattice at $\beta = 2.5115$. The maximal Abelian
gauge fixing used over-relaxation.   Fig.(2) shows the profile of the electric
flux corresponding to $\widehat{F}^{(2)}$ and $\widehat{F}^{(3)}$.
Fig.(3) shows the profile of the theta component of the magnetic current
corresponding to $\widehat{F}^{(1)}$, $\widehat{F}^{(2)}$ and $\widehat{F}^{(3)}$.

In summary, we showed that consistency requires one use the same definition of flux
throughout.  If, for example, one uses  $\widehat{F}^{(3)}$ definition of electric field (bottom
graph in Fig.(2)) in order to account correctly for the total flux but then uses the
DT definition of current (top graph in Fig.(3)) we would then incur errors 
of $\sim 40$\%.  For simulations in the scaling window, as long as we use the same
definition consistently we do not foresee 
large changes in the  standard analysis of the dual Abrikosov vortex
in the maximal Abelian gauge because the order $a^2$ corrections have small fluctuations and tend to
cancel out\cite{poulis}.  However in other gauges, the consequences of our definitions could lead to large effects 
which may help in understanding the choice of gauge.


\includegraphics[trim=0 52mm 0 47mm,scale=0.35]{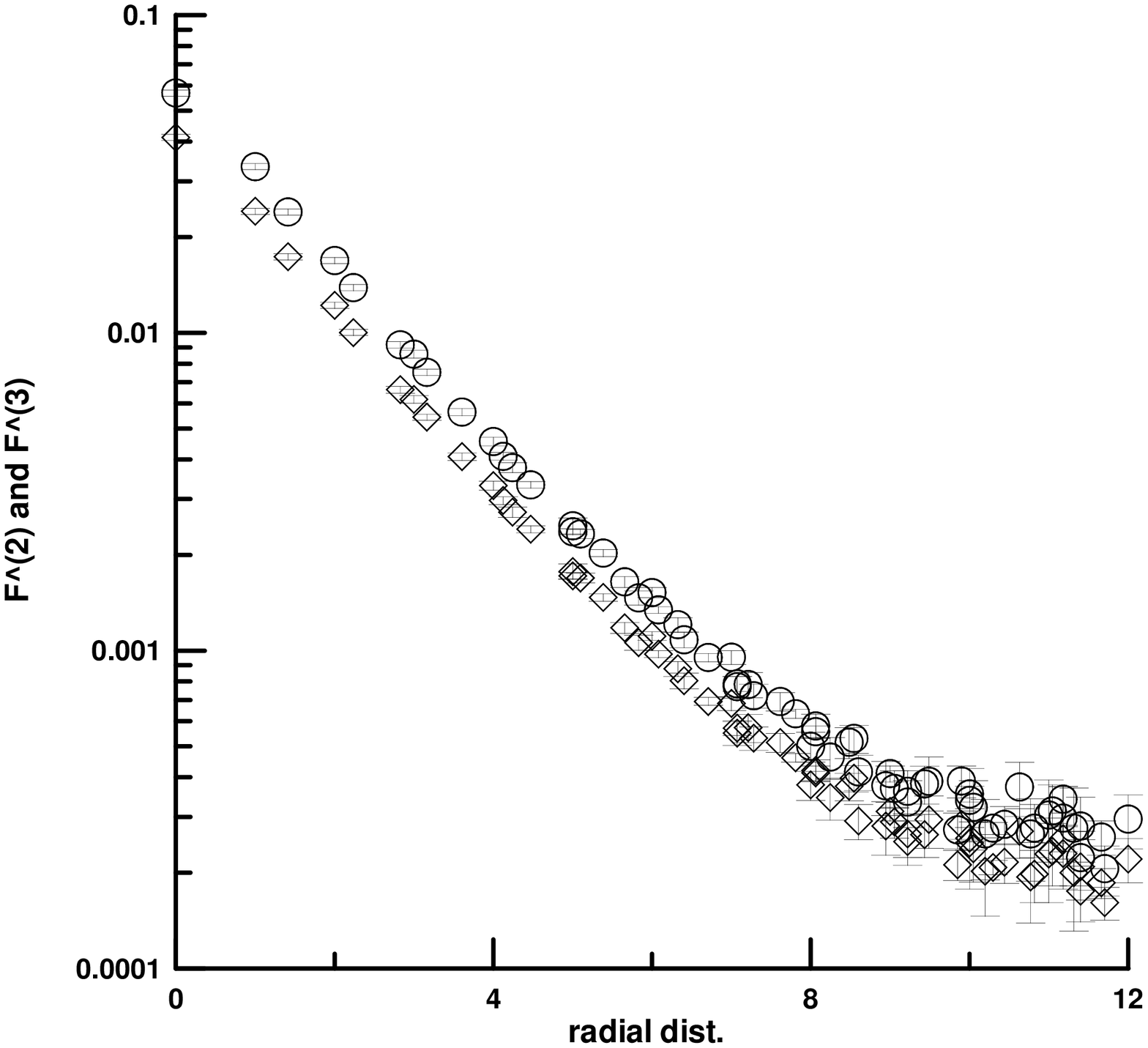}
\label{mars}

\begin{flushleft}
Figure 2. Profile of the electric field (highest to lowest) 
$\widehat{E}_3^{(2)}$ (circles) and $\widehat{E}_3^{(3)}$ (diamonds)
on the mid-plane 
between $q$ and $\bar{q}$ separated by $13a$.
\end{flushleft}


\includegraphics[trim=0 47mm 0 47mm,scale=0.35]{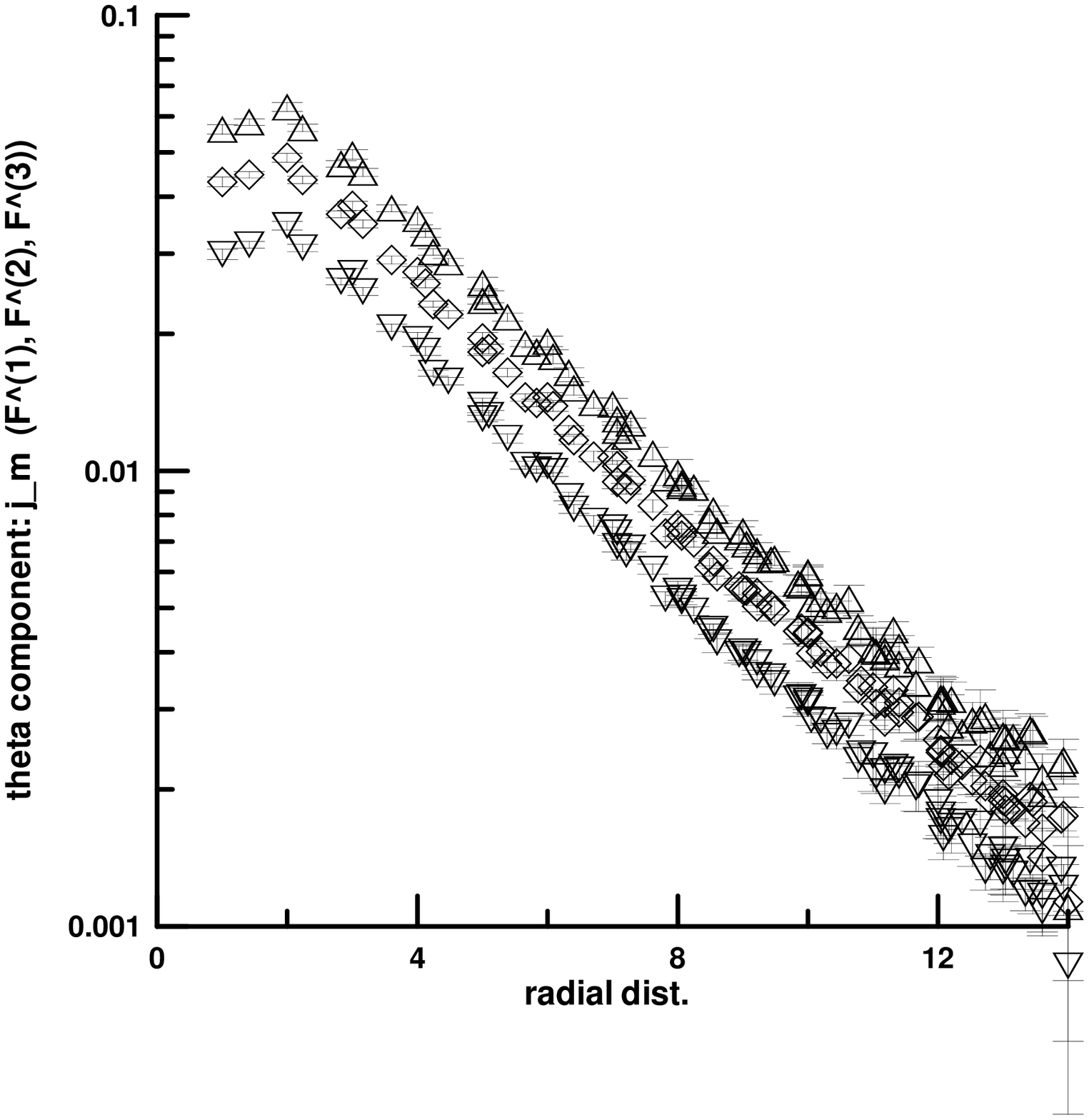}
\label{jupiter}

\begin{flushleft}
Figure 3. Profile of the theta component of the magnetic current on the
mid-plane between $q$ and $\bar{q}$ separated by $13a$
based on (highest to lowest) $\widehat{F}_{\mu\nu}^{(1)}$ (triangles),
$\widehat{F}_{\mu\nu}^{(2)}$ (diamonds) and  $\widehat{F}_{\mu\nu}^{(3)}$
(inverted triangles).
\end{flushleft}

Finally we report on the effect of truncating DT monopole loops, keeping only the one large
connected cluster\cite{ht,z}.  As mentioned in the Introduction 
this truncation is expected to have no effect on the confinement
signal.   This should manifest itself here in that the tail of the profile
of the electric field and magnetic current of the vortex are unaffected by the truncation.
This is born out as expected.  Fig.(4) shows that radial profile of the magnetic current
is indistiguishable except for a small deviation in the core of the vortex.


\includegraphics[trim=0 48mm 0 52mm,scale=0.33]{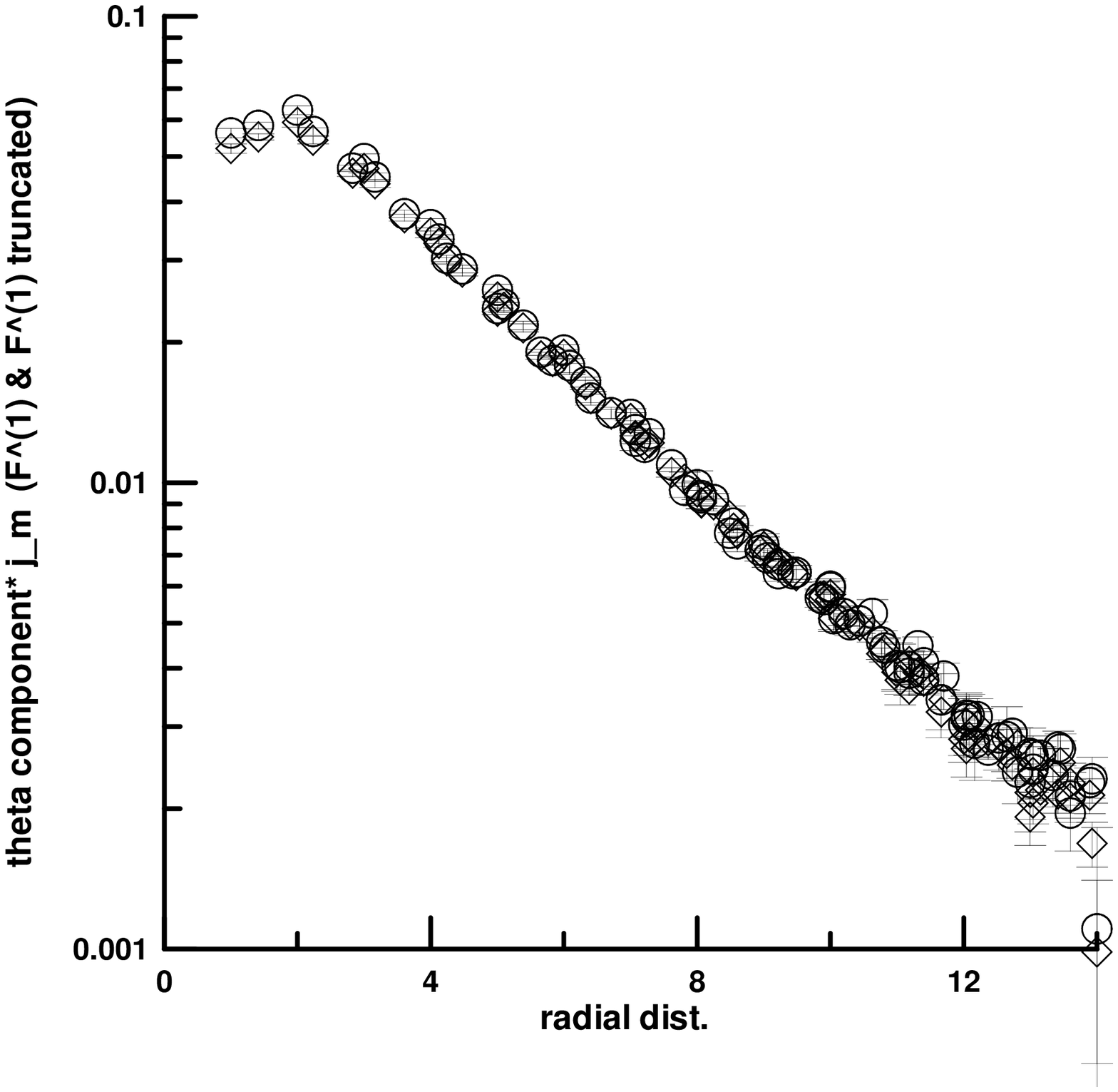}
\label{saturn}

\begin{flushleft}
Figure \ref{saturn}. Profile of the theta component of the magnetic current on the mid-plane between 
$q$ and $\bar{q}$ separated by $13a$
 based on (highest to lowest) $\widehat{F}_{\mu\nu}^{(1)}$ (circles),
 $\widehat{F}_{\mu\nu}^{(1)}$ truncated (diamonds).
\end{flushleft}

\section*{Acknowledgments}
This work is supported in part by the U. S. Department of 
Energy under grant no. De-FG05-01 ER 40617.

\end{document}